\documentclass{natureprintstyle}

\usepackage{amssymb}
\usepackage{graphicx,subfigure,color}
\usepackage{pdfpages}

\def\kms{km s$^{-1}$}

\def\arcsec{$^{\prime \prime}$}
\newcommand{\farc}{\mbox{\ensuremath{.\!\!^{\prime\prime}}}}
\definecolor{Mygrey}{gray}{0.75}

\newcommand{\ltsimeq}{\raisebox{-0.6ex}{$\,\stackrel{\raisebox{-.2ex}{$\textstyle <$}}{\sim}\,$}}

\title{A black-hole mass measurement from molecular gas kinematics in NGC4526}
\author{Timothy A. Davis$^{1}$, Martin Bureau$^2$, Michele Cappellari$^2$, Marc Sarzi$^3$ \& Leo Blitz$^4$}

\begin{document}

\maketitle

\begin{abstract}
The masses of the supermassive black-holes found in galaxy bulges are correlated with a multitude of galaxy 
properties\cite{magorrian,graham}, leading to suggestions that galaxies and black-holes may evolve together\cite{silkrees}.
The number of reliably measured black-hole masses is small, and the number of methods for measuring them is limited\cite{FandF}, holding back attempts to understand this co-evolution. Directly measuring black-hole masses is currently possible with stellar kinematics (in early-type galaxies), ionised-gas kinematics (in some spiral and early-type galaxies\cite{Sarzi01,Barth01,Ho2002}) and in rare objects which have central maser emission\cite{low}. Here we report that by modelling the effect of a black-hole on the kinematics of molecular gas it is possible to fit interferometric observations of CO emission and thereby accurately estimate black hole masses. We study the dynamics of the gas in the early-type galaxy NGC~4526, and obtain a best fit which requires the presence of a central dark-object of 4.5$^{+4.2}_{-3.0}$$\times$10$^8$~M$_{\odot}$ (3$\sigma$ confidence limit). With next generation mm-interferometers (e.g. ALMA) these observations could be reproduced in galaxies out to 75 megaparsecs in less the 5 hours of observing time. The use of molecular gas as a kinematic tracer should thus allow one to estimate black-hole masses in hundreds of galaxies in the local universe, many more than accessible with current techniques.
\end{abstract}
\begin{figure*}
\begin{center}
\includegraphics[width=0.85\textwidth]{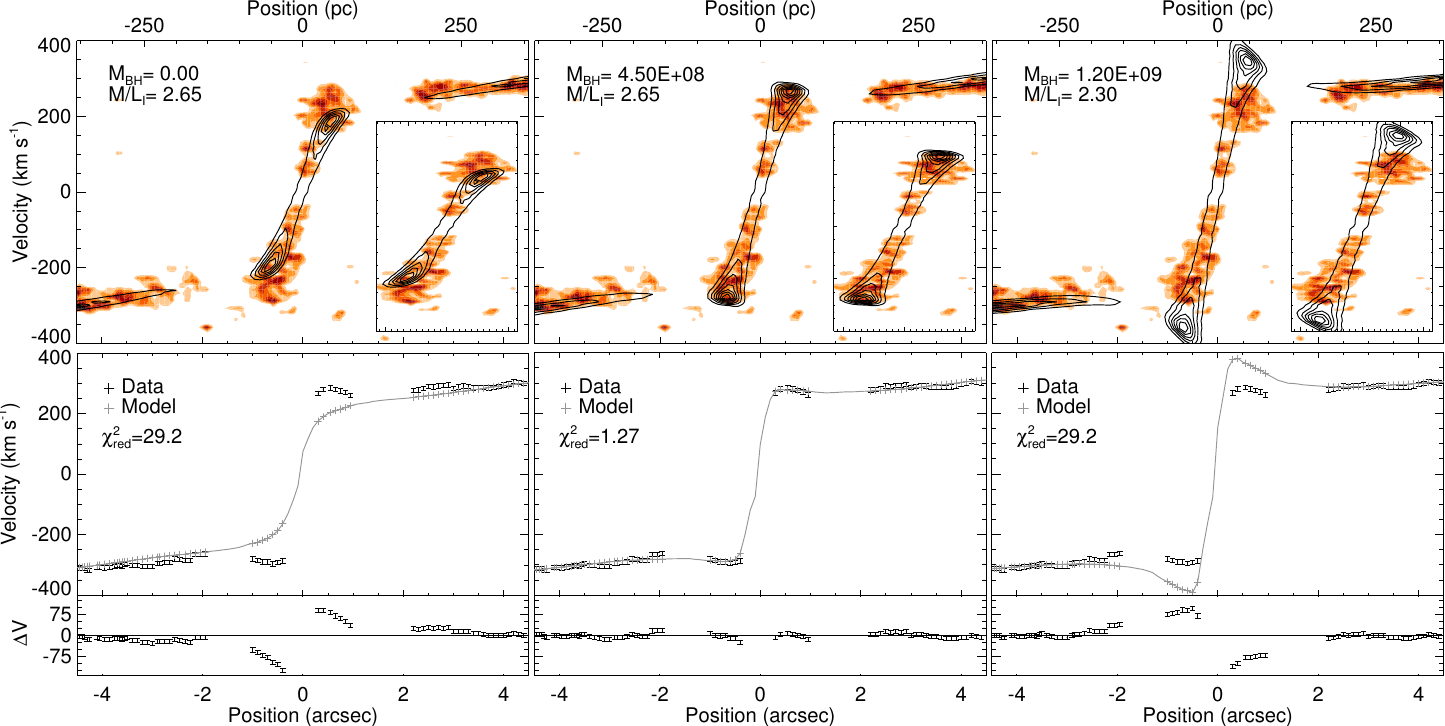}
\caption{{\bf NGC~4526 kinematic models and data.} Top: Model PVDs (black contours),
      overlaid on the observed CO(2-1) PVD (filled orange contours).
This PVD was created from our CO(2-1) observations of NGC~4526 from CARMA. The synthesized beam size achieved in these observations is $0\farc27\times0\farc17$, and the velocity channel width is $10$~\kms. The final fully reduced and calibrated data cube has an RMS noise of $2.88$~mJy~beam$^{-1}$.  The PVD was created by rotating the data cube to align the
  kinematic major axis of the molecular gas with the $x$-axis, and
  then summing over one beam width around the axis in
  the $y$ direction. The spatial resolution achieved in the PVD is $0\farc25$ ($20$~pc), equal to the predicted SMBH SOI. Our results do
not depend on the method used to extract the PVD. From left-to-right, the best model with no
      SMBH, the overall best-fit model, and a model with an overweight
      SMBH. The model $M_{\rm BH}$ and $M/L_I$ are indicated in the
      top-left corner of each panel, and an inset of the central
      $\pm1\farc15$ is shown in the bottom-right corner. 
      Middle: Black points show the trace extracted from the observed PVD, and associated standard errors. The grey-line shows the trace extracted from the models (and the grey crosses denote the value of the trace at the same radius as the observed points). Bottom: Residuals between the model and data at each position ($\Delta$V=data minus model; \kms). The error bars shown in the middle and bottom plots correspond to the formal uncertainties in fitting the trace (see Section 1.2 in the supplementary information), added in quadrature with two factors of $5$~\kms\ (to account for the finite velocity channel width in both the data and model).
\label{fig2}}
\end{center}
\end{figure*}

For this study, we 
targeted the fast-rotating early-type galaxy (ETG) NGC~4526. This object has a stellar velocity dispersion within one
  effective radius\cite{sauron4} $\sigma_{\rm e}=222$~\kms\ . 
This galaxy has not had its black-hole mass measured by any other technique to date
   but, given that most galaxies with bulges seem to have a black hole\cite{graham,Sarzi01},
   from the  $\sigma-M_{\rm BH}$ relation\cite{gult} we can estimate 
 $M_{\rm BH}\approx2\times10^8$~$M_{\odot}$.
 The SMBH sphere
  of influence (SOI) would thus be $r_{\rm SOI}=GM_{\rm
    BH}/\sigma^2\approx20$~pc or $0\farc25$ at the $16.4$~Mpc
  distance of NGC~4526 (derived from surface-brightness fluctuations\cite{tonry}). 
    NGC~4526 also has a large molecular gas reservoir (total molecular hydrogen
  mass $M_{{\rm H}_2}=3.69\times10^8$~$M_{\odot}$\cite{Combes}),
  previously mapped at low spatial resolution\cite{YBC} ($\approx$4\arcsec).
The molecular gas co-rotates with the stars\cite{davis_origin} and Hubble
  Space Telescope (HST) images show that it is coincident with regular
  dust lanes extending up to the galaxy centre,
  strongly suggesting that molecular gas with regular kinematics
  exists around the SMBH\cite{Ho2002}.

  \let\thefootnote\relax\footnote{
\begin{affiliations}
 \item European Southern Observatory, Karl-Schwarzschild-Str. 2, 85748, Garching-bei-M\"unchen, DE
 \item Sub-department of Astrophysics, Department of Physics, University of Oxford, Denys Wilkinson Building, Keble Road, Oxford, OX1 3RH, UK
  \item Centre for Astrophysics Research, University of Hertfordshire, Hatfield, Herts AL1 9AB, UK
   \item Department of Astronomy, University of California, Berkeley, CA 94720, USA
\end{affiliations}
}

NGC~4526 was observed in the CO(2-1) line ($230$~GHz)  using the Combined Array for Research in Millimetre Astronomy (CARMA) in A, B and C configurations\cite{bock}. The data were reduced in the standard manner\cite{A12,miriad}, and more details are presented in Figure 1. The spatial resolution achieved along the kinematic major axis of the galaxy is $0\farc25$ ($20$~pc), equal to the predicted SMBH SOI. The gas in this source seems to be distributed in a central component, and an inner ring with spiral spurs which lead outwards to another ring at larger radii. A detailed analysis of the molecular gas morphology will be conducted in a future work. We note, however, that the gas structures lie in the plane of the galaxy and appear to be regularly rotating; they are therefore likely to be dynamically cold, and a good tracer of the potential\cite{davis_morphkin}.

We created a grid of simulations of NGC~4526 spanning a range of SMBH
masses (no SMBH, $M_{\rm BH}$= 3$\times10^6$, and then 5$\times10^7$ -- $1.45\times10^{9}$~$M_\odot$ in linear steps) and $I$-band mass-to-light ratios ($M/L_I=0.55$ -- $6.15$~$M_\odot/L_\odot$ in linear steps). For full details of these simulations see Section 1.1 in the online supplementary information. We fix the inclination of the gas disk\cite{davis_tfr} ($i=79^\circ$), and use an axisymmetric mass model of NGC~4526\cite{sauron4} (carefully fitted to avoid contamination due to dust; see Section 1.1.2 in the supplementary material) to derive the circular-velocity curve expected from the luminous matter alone. 
The presence of a SMBH in NGC~4526 manifests
itself as an inner Keplerian rise of the rotation curve (above that expected
from luminous matter only). On larger angular scales such fast-rising rotation curves have been observed, and used to infer the masses of central star clusters and bulges\cite{sofue}.
We fit these models to our observed data in order to determine if such an excess due to a central dark mass is detectable in NGC~4526.

Figure~\ref{fig2} shows three different simulated position-velocity diagrams (PVDs) overlaid on the
observed PVD of NGC~4526, with $\pm1\farc15$ insets. The first panel
shows the best model with no SMBH,
with a clear excess of high-velocity molecular gas at the centre. The
middle panel shows our overall best-fit model, clearly reproducing
better the observed PVD at all radii, with
$M_{\rm BH}=4.5\times10^8$~$M_\odot$ and
$M/L_I=2.65$~$M_\odot/L_\odot$. The final panel shows a model with a
much larger SMBH, clearly incompatible with our data. The lower panels of Figure~\ref{fig2} show the trace extracted from the observed PVD (as described in the supplementary material), and associated errors. The grey line shows the trace extracted from the models in the same way (with points denoting the values at the same radius as the observed points). The residuals between the data and the model at each position ($\Delta$V=data minus model) are shown in the bottom panel. Clearly the best-fit model produces a significantly better fit to the data.

Figure~\ref{fig3} shows the $\chi^2$ contours of our
fits (conducted as described in the supplementary material) as a function of $M_{\rm BH}$ and $M/L_I$. A clear global minimum
is present at the best-fit values, with a minimum reduced $\chi^2$ of 1.27. 
 We define uncertainties for each of our fitted parameters from likelihood functions of each
parameter (marginalised over the other), by finding the region containing respectively $68\%$ and $99\%$
of the probability. Our final best-fit values are
${M_{\rm BH}=4.5^{+1.5}_{-1.3}\times10^8}$~${M_\odot}$ with
$M/L_I=2.65{\pm0.21}$~$M_\odot/L_\odot$ ($68\%$ confidence
level), and 
$4.5^{+4.2}_{-3.1}\times10^8$~${M_\odot}$ with
$M/L_I=2.65^{+0.56}_{-0.52}$~$M_\odot/L_\odot$ ($99\%$ confidence
level). These values are fully
consistent within the 3$\sigma$-uncertainties with the reported $\sigma_{\rm
  e}$-$M_{\rm BH}$ relation\cite{gult}. Our formal uncertainties (0.14 dex and 0.5 dex at 1$\sigma$ and 3$\sigma$, respectively) are also similar (and in many cases smaller) to the
average uncertainties reported by other authors. For instance the mean 1$\sigma$ error reported when using dynamical stellar and ionised gas techniques is $\approx$0.6 dex\cite{gult} (however the rather different systematic errors involved make direct comparison difficult).

{ The $\chi^2$ contours show the usual degeneracy between
  SMBH mass and mass-to-light ratio found in other studies.
  To fit the data without a SMBH, a negative $M/L$ gradient would be
  necessary, with older stellar populations dominating in the inner
  parts (high $M/L$) and young stars at large radii (low
  $M/L$). While the real $M/L$ is unlikely to be exactly
  constant, this is opposite to the trend reported by reddening free stellar
  population studies\cite{sauron17} and that expected from the
  presence of molecular gas and star-formation in the central regions\cite{ali_sf}.}

The ability to determine a black-hole mass accurately using molecular-gas can be affected by many of the same issues that affect measurements of ionised-gas. Turbulent motions are in general small in molecular gas, but could conceivably increase around a black-hole. Similarly, if the inner gas were to be misaligned from the stellar body, our mass estimate would be systematically affected in a way that is degenerate with a change in $M/L$.  In this galaxy, however, we find no evidence that the velocity dispersion increases in the inner regions (as described in Section 1.1.1 in the supplementary material), and constrain the inner gas to be aligned with the stellar body within $<$3$^{\circ}$. Such a misalignment could change velocities by an insignificant amount ($\ltsimeq$3\kms; see Section 1.1.2  in the supplementary material). The presence of dust could also cause mass models to underestimate the contribution of luminous matter to the potential. We again believe this should not introduce significant errors in this object, because of our careful treatment of dust in the mass model (see Section 1.1.2 in the Supplementary Information). Future studies using this technique should choose their targets to minimise the impact of such effects, but can also include warps and turbulent motions in their gas disk models\cite{neumayer}, and use near-infrared photometry as done in previous studies of dusty objects\cite{neumayer}. 

The use of molecular gas as a kinematic tracer holds great promise to
increase the total number of SMBH mass measurements in galaxies of all
types where CO is detected. 
Angular resolutions of up to $0\farc15$ can be
achieved with current mm-interferometers, allowing for instance to
resolve the SOI of a $2\times10^8$~$M_\odot$ SMBH in a galaxy with
$\sigma_{\rm e}=200$~\kms\ only out to around $30$~Mpc.
The next generation of millimetre interferometers will have over an order of magnitude greater angular resolution (for example, $\approx6$~mas at $690$~GHz with ALMA), as well as greatly increased sensitivity. 
In less the 5 hours of integration time with ALMA one could achieve the same sensitivity and linear resolution as the observations presented here, in a galaxy 75 Mpc away.
Galaxies of lower mass (with smaller SMBHs) will also become accessible. 
For instance, the SOI of a Milky Way-like SMBH\cite{ghez}
($M_{\rm BH}=4\times10^6$~$M_\odot$, $\sigma_{\rm e}=105$~\kms) will be resolved up to $\approx50$~Mpc. 

Measurements using a single technique will thus be
possible over the entire range of the $\sigma_{\rm e}$-$M_{\rm BH}$
relation, leading to much needed reduced uncertainties in the slope
and normalisation of black hole-galaxy relations.
This will allow the study of a large number of spiral
galaxies, that cannot currently be probed easily using stellar
dynamical techniques. Furthermore, it will also provide access to a larger number of systems than ionised gas techniques (currently limited by the $\approx$0\farc05 resolution of HST).
Even considering that only some of the accessible objects will have
suitably relaxed and centrally-peaked distributions of molecular gas, measuring SMBH masses will be possible in
many hundreds of spiral (and early-type) galaxies, many times more than possible using
conventional stellar and ionised-gas tracers today. 
This technique could also be extended using other spectral lines, such as the very bright atomic gas cooling lines at higher frequencies (in order to increase the spatial resolution achieved even further), and optically thin transitions such as $^{13}$CO (which would allow determinations of black-hole masses even in systems that are exactly edge on).

\begin{figure}
\includegraphics[width=0.5\textwidth,clip,trim=1.7cm 0.3cm 0.9cm 1.0cm]{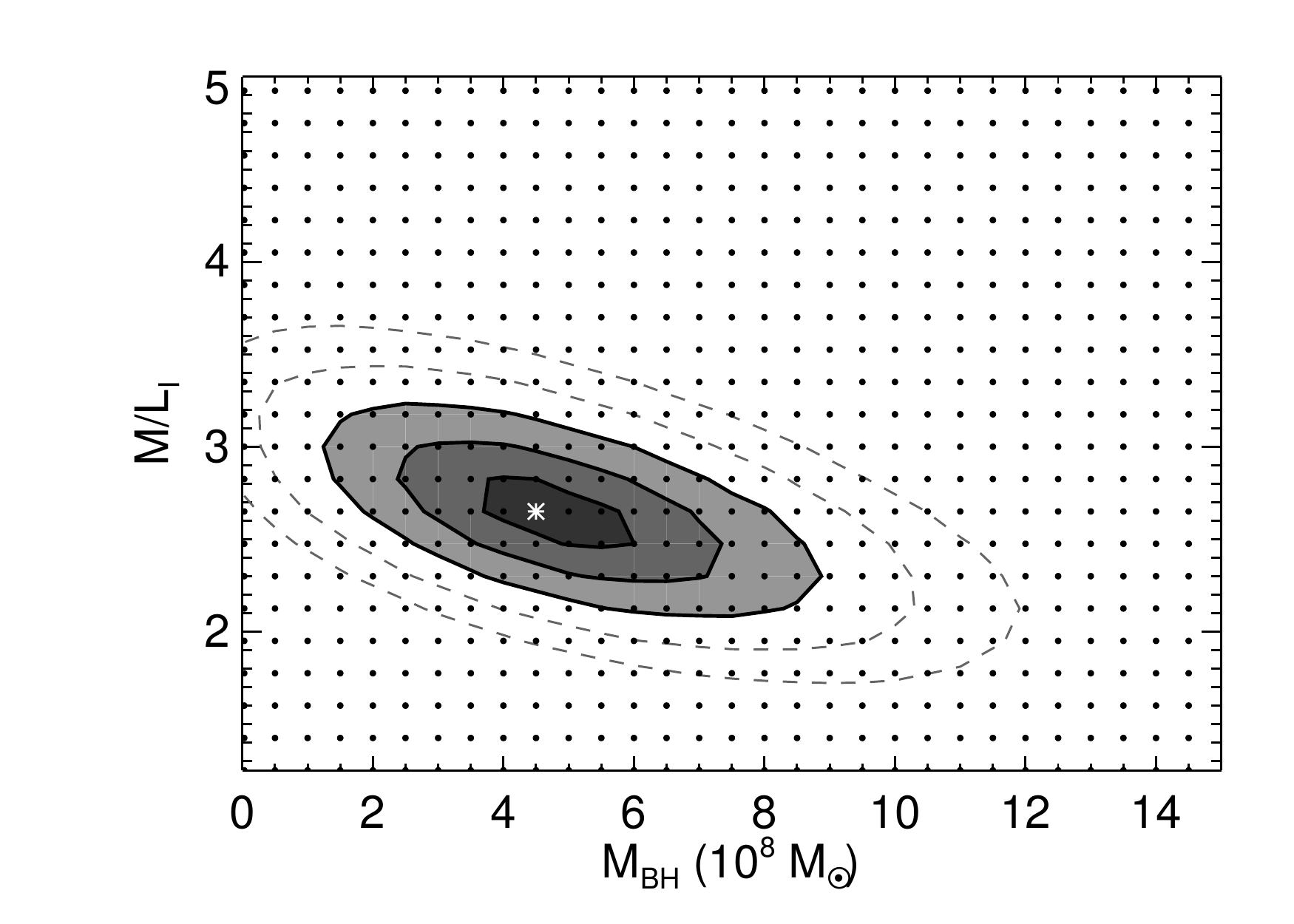}
\caption{{\bf NGC~4526 SMBH mass uncertainties.} 
    $\Delta\chi^2\equiv\chi^2-\chi_{\rm minimum}^2$ contours of our fits to the CO(2-1) PVD, as a function
    of the two free parameters $M_{\rm BH}$ and $M/L_I$. The model
    grid is shown with black dots and the overall best-fit model by a
    white star. The solid shaded contours with black lines correspond to the 1-3$\sigma$ levels with one degree-of-freedom ($\Delta\chi^2$= 1, 4, 9) with good models in the darkest
    areas. The 4 \& 5$\sigma$ levels ($\Delta\chi^2$= 16, 25) are denoted with dashed grey lines.\label{fig3}}
\end{figure}

\noindent
\textbf{\textsf{\footnotesize Received 11th September; accepted 27th November 2012}}

\begin{addendum}
\item The research leading to these results has received funding from the
European Community's Seventh Framework Programme. MB is supported by the rolling grants
`Astrophysics at Oxford' and from the UK
Research Councils. MC acknowledges support from a Royal Society
University Research Fellowship. MS acknowledges support from a Science and Technology Facilities Council (STFC)
Advanced Fellowship.  Support for CARMA construction was
derived from the states of California, Illinois, and Maryland, the
James S.\ McDonnell Foundation, the Gordon and Betty Moore Foundation,
the Kenneth T.\ and Eileen L.\ Norris Foundation, the University of
Chicago, the Associates of the California Institute of Technology, and
the National Science Foundation. Ongoing CARMA development and
operations are supported by the National Science Foundation under a
cooperative agreement, and by the CARMA partner universities.
 \item[Author Contributions]  T.A.D prepared and reduced the observations, and created the modelling tool. TAD and MB prepared the manuscript. MC created the mass model. All authors discussed the results and implications and commented on the manuscript at all stages.
 \item[Author Information] Reprints and permissions information is available at
www.nature.com/reprints. The authors declare no competing financial interests.
Readers are welcome to comment on the online version of this article.
Correspondence and requests for materials
should be addressed to T.A.D~(email: tdavis@eso.org).

\end{addendum}

\clearpage


\section*{Supplementary material}

# Supplementary Information

## 1.1 Model Observations

In this paper we compare the observed CO position-velocity diagram with one created by modelling the gas disk. When creating these models we included the observational effects of beam-smearing, disk-thickness, velocity dispersion, binning, etc, by utilising the KINematic Molecular Simulation (KinMS) mm-wave observation simulation tool[16,25]. The simulations used a beam, pixel size and velocity resolution identical to our observations, and an idealised surface brightness profile which fits the data. We use an axisymmetric mass model of NGC 4526 to derive the circular-velocity curve expected from the luminous matter alone. These elements of the models are described in detail below.

### 1.1.1 Gas distribution

The distribution of the molecular gas in this source is a required input for the KinMS tool. As we primarily aim to assess the kinematic agreement between the data and models with different SMBH masses (rather than the details of the gas distribution itself), a simple surface brightness profile fitting the observed distribution is sufficient. We therefore adopted a profile composed of a central Gaussian with full-width-at-half-maximum FWHM=$0\rlap.''15 \approx$ 12 pc, centred at R=$0.6'' \approx$ 48 pc (nuclear ring) and a Gaussian of FWHM=$2'' \approx$ 160 pc centred at R=$5'' \approx$ 400 pc (main ring). This parameterized surface brightness profile fits well the radial distribution of the observed clean components within the inner $6''$, and identical results are obtained for all sensible profile choices. We assumed a gas velocity dispersion of 8 km s$^{-1}$ here (our results do not depend on choosing any reasonable value for this parameter). Models with $\sigma_{\rm gas} <$ 10 km s$^{-1}$ are required to match the observed molecular gas distribution at all radii.

### 1.1.2 Mass Model

The mass model we use in this work is based on a multi-Gaussian expansion (MGE[26]) of F814W HST and MDM 1.3-m telescope $I$-band images (probing both small and large scales), and is parametrised by the $I$-band stellar mass-to-light ratio $M/L_I$ and inclination $i$. From it the circular velocity of the galaxy can be computed at any radius. The model is carefully fitted to remove any contamination from dust still visible in the $I$-band. The dust disk in this system is well defined, and we have many resolution elements inside the SOI of the SMBH. This allows us to carefully mask the affected regions when conducting the MGE fit[27]. The unaffected sectors were then used to derive the MGE model. Such a process has been shown to work well at recovering the intrinsic light distribution in such systems[6].

Axisymmetric two-integral Jeans and three-integral Schwarzschild dynamical models of the large-scale stellar kinematics (within $45''$) suggest $M/L_I =$ 3.35 $M_\odot/L_\odot$ and $i = 79°$, but stellar population models of the inner parts ($\lesssim 5''$) suggest $M/L_I =$ 2.6 $M_\odot/L_\odot$ (derived assuming a Kroupa IMF[6], which is likely appropriate for this galaxy[28]), presumably because of young stellar populations associated with the molecular gas. Any residual dust obscuration in the very centre of the galaxy could cause us to underestimate the M/L in the dynamical model, however the stellar population $M/L_I$ estimate (derived from an analysis of line-strength indices close together in wavelength[18]) is dust-independent.

The inclination estimated from the dynamical models is consistent with that derived for the dust disk and the spread of measured values is very small[17] ($\pm 3°$).

Ellipse fitting to the inner dust lanes visible in HST imaging, and to the inner molecular component show that the inner gas is also aligned with the stellar body within $<3°$ As NGC 4526 is close to edge on, such an uncertainty in inclination produces an insignificant error in the projected velocities ($\approx$ 3 km s$^{-1}$), which is insignificant when compared to the uncertainties introduced by other parameters (see Section 1.2 of the supplementary material). We therefore fix the inclination of the gas disk at all radii to $i = 79°$.

## 1.2 Trace extraction

For a quantitative comparison focusing on the kinematics, we extracted the 'trace' of the outer envelope of the observed and simulated PVDs, every $0\rlap.''1$ in radius (in order to Nyquist sample our beam), by fitting a Gaussian (along the velocity axis) to each spatial bin and taking the central velocity of the Gaussian plus one standard deviation as the trace. The uncertainties were estimated from the formal uncertainties in the velocity and standard deviation of the fitted Gaussian, plus two factors of 5 km s$^{-1}$ (to account for the finite velocity channel width in both the data and model). These channel width factors usually dominate.

## 1.3 Fitting procedure

The best fit was found using a standard $\chi^2$ test, comparing the one-dimensional trace of the data and simulations over the central $\pm 4\rlap.''5$, the area most affected by a putative SMBH (See Figure 2). We did not fit inside the innermost beam, where the effects of beam smearing are strongest, leaving us with 68 observed data-points. The best-fit profile has a $\chi^2$=86.36, equivalent to a reduced $\chi^2$=1.27.

4



\end{document}